 \documentclass[12pt,preprint]{aastex}

\usepackage{amsmath,amssymb}

\received{: September 9, 2010}
\accepted{: October 15,2010}

\slugcomment{Published in {\it Astropysics and Space Science},  {\bf  332},  365--371   (2011)}

\shorttitle{Approximation for Chandrasekhar's $H$-function}
\shortauthors{Kawabara \& Limaye}

\begin{document}


\title{Rational Approximation Formula \\ for Chandrasekhar's $H$-function \\ for Isotropic Scattering}


\author{Kiyoshi Kawabata}
\affil{Department of Physics, Tokyo University of Science,
  Shinjuku-ku, \\ Tokyo 162-8601, Japan}
\email{kawabata@rs.kagu.tus.ac.jp}
\author{Sanjay S. Limaye}
\affil{Space Science and Engineering Center, University of Wisconsin,
 Madison, \\ Wisconsin 53706, USA}
\email{SanjayL@ssec.wisc.edu}



\begin{abstract}
In this work, we first establish a simple procedure to obtain with 11-figure accuracy the values of Chandrasekhar's $H$-function for isotropic scattering using a closed-form integral representation and the Gauss-Legendre quadrature.
Based on the numerical values of the function produced by this method for various combinations of $\varpi_0$, the single scattering albedo, and $\mu$, the cosine of the zenith angle $\theta$ of the direction of radiation emergent from or incident upon a semi-infinite scattering-absorbing medium, we propose a rational approximation formula with $\mu^{1/4}$ and $\sqrt{1-\varpi_0}$ as the independent variables. This allows us to reproduce the correct values of $H(\varpi_0,\mu)$ within a relative error of $2.1\times 10^{-5}$ without recourse to any iterative procedure or root-finding process.  
\end{abstract}




\section{Introduction}
Chandrasekhar's H-functions $H(\varpi_0,\mu)$ are used to express the emergent intensities of radiation reflected by semi-infinite, homogeneous media. They satisfy the following nonlinear integral equation \citep[see, e.g., ][]{chan1960}:
\begin{equation}
\displaystyle{H(\varpi_0,\mu)=1+\mu H(\varpi_0,\mu)\int_0^1{\Psi(\eta)\over \mu+\eta}H(\varpi_0,\eta)d\eta},  \label{eq-for-H}
\end{equation}
where $\varpi_0$ is the single scattering albedo, and $\Psi$, which is also a function of $\varpi_0$, is the characteristic function specific to the type of scattering the radiation undergoes. 
We developed, for instance, a compact computational method to find the numerical solutions of Eq.(\ref{eq-for-H}) 
for some representative types of anisotropic scattering in terms of the roots of the characteristic equation involving values of $\Psi$ evaluated at the quadrature points and the associated weights of integration employed \citep[][]{kaw1991}.\par
In practical applications of the theory of radiative transfer, however, we often need a quick and yet relatively accurate approximation for the $H$-function, especially for isotropic scattering. In the present study, we therefore focus on developing a simple numerical procedure to calculate accurately (at least to the 11th digit) the values of $H(\varpi_0,\mu)$ for conservative as well as non-conservative isotropic scattering.
We will then construct a rational approximation formula 
based on fitting to the reference values of $H(\varpi_0,\mu)$ produced with the foregoing method. This permits us to avoid the iterative methods that are often employed in this type of computations. \par
We must note that several useful approximation formulae have been proposed by various authors. The formula given by \citet{kar1971} for arbitrary values of $\varpi_0$ is sufficiently compact,
but its use requires the solution of a transcendental equation for a given value of $\varpi_0$, which we want to avoid. 
The same is true for the formula derived by \citet{dom1988}. 
The formula proposed by \citet{hap1993} is also handy, but its accuracy is around $7.7\times 10^{-3}$. 
The formula developed by \citet{kar1991} is sufficiently accurate, with a maximum relative error of about $10^{-4}$, and requires no root-finding process, but is valid only for conservative scattering. Recently, however, \citet{dav2008} obtained a new formula for an arbitrary set of $(\varpi_0,\mu)$, which is an order of magnitude more accurate (the maximum relative error is $7\times 10^{-4}$) than any of these. 
Therefore, it must be an interesting challenge to find a formula that is at least an order of magnitude better than this. We will make all the numerical computations required for this objective exclusively in double-precision arithmetic.
\section{Formulations and Numerical Computations}
\subsection{Integral Representation}
The H-function for isotropic scattering $H(\varpi_0,\mu)$ can be expressed in a closed-form integral representation \citep[see ][]{Rut1987}:
 \begin{equation}
 \displaystyle{H(\varpi_0,\mu)=\exp\left[-{\mu\over \pi}\int_0^\infty \ln\,(1-\varpi_0 \xi \ {\rm arccot}\,\xi)\cdot{d\xi\over \mu^2+\xi^2}\right]}. \label{eq-1}
\end{equation}
Substituting ${\rm arccot}\, \xi=x$ in Eq.(\ref{eq-1}), we have
\begin{equation}
H(\varpi_0,\mu)=\displaystyle{\exp\left[-{\mu\over \pi}\int_0^{\pi/2}\ln(1-\varpi_0\,x\cot\,x)\cdot{(1+\cot^2x)\over (\mu^2+\cot^2x)}dx\right]}. \label{eq-Hx} 
\end{equation}\par
Eq.(\ref{eq-Hx}) is advantageous for numerical
 computations because the domain of integration is finite. We would therefore like to use this expression along with the Gauss-Legendre quadrature to generate the reference values of $H(\varpi_0,\mu)$ for the purpose of comparison. 
For non-conservative scattering, for which $\varpi_0 < 1$, the numerical integration inside Eq.(\ref{eq-Hx}) can be carried out without difficulty by means of the quadrature, and the resulting values of 
$H(\varpi_0,\mu)$ should be accurate enough even if a single quadrature is applied to the entire domain of integration $[0, \pi/2]$.
However, conservative scattering with $\varpi_0=1$ poses a numerical problem in that the factor $1-x\cot x$, inside the natural logarithm involved in the integrand, diverges as $x$ tends to $0$,
which could significantly degrade the numerical accuracy of the resulting H-function. With this in mind, for conservative scattering, we divide 
 the domain of integration into two parts, $[0, \varepsilon]$ and $[\varepsilon, \pi/2]$, where $\varepsilon \ll 1$:
\begin{equation}
\displaystyle{\int_0^{\pi/2}\ln(1-x\cot\,x)\cdot{(1+\cot^2x)\over (\mu^2+\cot^2x)}dx=I_1+I_2},
\end{equation}
where 
\begin{equation}
\displaystyle{I_1\equiv \int_0^\varepsilon\ln(1-x\cot\,x)\cdot{(1+\cot^2x)\over (\mu^2+\cot^2x)}dx},     \label{eq-I1-integ}
\end{equation}
and
\begin{eqnarray}
I_2&\equiv& \displaystyle{\int_\varepsilon^{\pi/2}\ln(1-x\cot\,x)\cdot{(1+\cot^2x)\over (\mu^2+\cot^2x)}dx}  \nonumber \\
&\simeq&\displaystyle{\sum_{j=1}^N\ln(1-x_j\cot\,x_j)\cdot{(1+\cot^2x_j)\over (\mu^2+\cot^2x_j)}w_j}, \qquad (\varepsilon < x_j < {\pi\over  2}),  \label{eq-I2-integ}
\end{eqnarray}
where $x_j$ and $w_j$ are the $j$-th quadrature point for evaluating the integrand and the corresponding weight of the Gauss-Legendre quadrature applied to the $x$-interval of $[\varepsilon, \pi/2]$, respectively. \par
Since $0 \le x\le \varepsilon\ll 1$ for $I_1$, we may expand its integrand in a  series in $x$. Retaining the terms up to and including the order of $x^3$, we get
\begin{equation}
\ln (1-x\cot x)\cdot \displaystyle{{1+\cot^2 x\over \mu^2+\cot^2 x}}\simeq \displaystyle{\ln {x^2\over 3}+\left[{1\over 15}+(1-\mu^2)\ln {x^2\over 3}\right]x^2+o[x]^4}.  \label{eq-expand}
\end{equation}
Substitution of the right-hand side of Eq.(\ref{eq-expand}) into Eq.(\ref{eq-I1-integ}) yields 
\begin{eqnarray}
I_1&\simeq&\displaystyle{\int_0^\varepsilon\left\{\ln {x^2\over 3}+\left[{1\over 15}+(1-\mu^2)\ln {x^2\over 3}\right]x^2\right\}dx} \nonumber \\
&=&\displaystyle{{\varepsilon\over 45}\left\{30(3+A\varepsilon^2)\ln \varepsilon+[1-5A\cdot(2+3\ln 3)]\varepsilon^2-45(2+\ln 3)\right\} }\nonumber \\
& &+o[\varepsilon]^5      \label{eq-I1}
\end{eqnarray}
with $A\equiv 1-\mu^2$ (see also Appendix A for a higher-order approximation).   
Because we are interested in producing the numerical values of $H(1,\mu)$ with 11-figure accuracy, we have the constraint $\varepsilon\le 10^{-2}$.  Let us therefore set $\varepsilon=10^{-3}$ for $\varpi_0=1$ and $\varepsilon=0$ otherwise. The value of $H(\varpi_0,\mu)$ owing to the integral representation is then given by 
\begin{equation}
H_{\rm integ}(\varpi_0,\mu)=\displaystyle{\exp\left[-{\mu\over \pi}(I_1+I_2)\right]}.    \label{eq-Hinteg}
\end{equation}
\indent Note that a similar integral representation was employed by \citet{dav2008} to produce the reference numerical values of $H(\varpi_0,\mu)$. However, their domain of integration was $[0, \infty]$, which makes it necessary to introduce a special scheme of integration. In view of this, we believe that the present technique is much more straightforward and easier to handle \footnote{The anonymous referee kindly directed our attention to an  alternative integral representation \citep[see, e.g.,  Eq.(16) of ][]{das2007}
.  
Based on some numerical tests  of this formula, we were however led to the conclusion that  the use of Eq.(3) coupled with Eq.(4) is approximately two orders of magnitude faster to produce the values of $H(1,\mu)$ that are correct to the 10th decimal place. A brief discussion on this matter is  given in Appendix B.}.
\subsection{Computational Results with Integral Representation}
We varied the degree $N$ of the Gauss-Legendre quadrature to calculate the values of $H(\varpi_0,\mu)$ for all the combinations of 48 values of $\varpi_0$ (0.01 plus 47 values employed by \citet{dav2008} for their Table 1) and 21 values of $\mu$ (0 plus 20 values employed by \citet{dav2008} for their Table 1). Note that similar numerical tables are also
given by \citet{bos1983}, but for the combinations of $\varpi_0$ = 0.5, 0.7, 0.9, 0.99, 0.999, and 1 and for $\mu$ from 0 to 1 with a step of 0.1.  \par 
We found that $N=100$ is sufficient to bring our results for $H(\varpi_0,\mu)$ into complete agreement with those of \citet{dav2008} and \citet{bos1983} down to the 10th decimal place (11 figures altogether).
 On the other hand, even if the 300th-degree Gauss-Legendre quadrature is employed for the single interval $[0, \pi/2]$, the accuracy of the resulting value of the H-function for the conservative scattering is much lower: 
we obtain, e.g.,   $H(1,1)=2.9077901976$ instead of the reference value 2.9078105291. 
This indicates the effectiveness of Eq.(\ref{eq-Hinteg}). \par
Next, to further assess the quality of the present approximation, we examined $\alpha_0$, the zeroth moment of the $H$-function, which can be expressed in terms of the single scattering albedo $\varpi_0$ as
\begin{equation}
\displaystyle{\alpha_0\equiv\int_0^1H(\varpi_0,\mu)d\mu}=\displaystyle{{2\over \varpi_0}\left[1-\sqrt{1-\varpi_0}\right]} 
\end{equation}
for isotropic scattering \citep[][]{chan1960}.
Using the values of $H(\varpi_0,\mu)$ given by our present method, we performed the numerical integrations required to obtain $\alpha_0$, again using the $N_{\rm GL}$-th degree Gauss-Legendre quadrature:
\begin{equation}
\displaystyle{\alpha_0}\simeq\displaystyle{\sum_{j=1}^{N_{\rm GL}}H(\varpi_0,\mu_j)w_j}.
\end{equation}
\indent Unfortunately, however, with $N=100$, which was sufficient for computing $H(\varpi_0, \mu)$, we found it impossible to get the value of $\alpha_0$ correct to the 9th decimal place regardless of how we chose the value of $N_{\rm GL}$. 
For instance, the best value we could obtain was $\alpha_0=2.0000000019$ with $N_{\rm GL}=130$ for $\varpi_0=1$. To obtain values of $\alpha_0$ correct down to the 9th decimal place, thereby allowing for at most a unit difference at the 10th decimal place relative to the theoretical values $2(1-\sqrt{1-\varpi_0})/\varpi_0$,
we found it necessary to employ $N=300$ and $N_{\rm GL}\ge 270$. Allowing for an adequate margin, we therefore adopted $N_{\rm GL}=350$, with which we finally obtained most of the computed values of $\alpha_0$ correct to the 10th decimal place, e.g., $\alpha_0=2.0000000000$ for $\varpi_0=1$ and $1.4854314511$ for $\varpi_0=0.88$. The only exception was that
for $\varpi_0=$ 0.9996, 0.9995, 0.999, 0.995,  0.993, 0.965, 0.95, 0.93, and 0.75, the figures at the 10th decimal place were larger than the theoretical values by unity. We nevertheless concluded that our primary objective was accomplished to a sufficient degree at this stage. 
\subsection{Approximation Formulae}
Now that we have established a highly reliable means of obtaining the numerical values of $H(\varpi_0,\mu)$,
we proceed to the next stage and seek a fast and yet reasonably accurate approximation formula
for $H(\varpi_0,\mu)$. For this, we shall proceed in two steps: (1) try to construct a polynomial approximation formula for $H(1,\mu)$ accurate to at least five figures, and (2) develop a rational approximation formula for $H(\varpi_0,\mu)$ by using that  obtained in the step (1). 
\hspace{3cm}\phantom{AAAA}\\
\subsubsection{Approximate Formula for Conservative Scattering }
Using polynomials of various degrees $K_1$ of $\mu^{1/4}$, we made least square fittings to the reference values of $H_{\rm integ}(1, \mu)$ tabulated for 501 equally spaced values of $\mu$ between 0 and 1:
 \begin{equation}
H_{\rm integ}(1,\mu)=\sum_{k=0}^{K_1} A_k x^k,
\end{equation}  
where $x=\mu^{1/4}$, as mentioned above, and $A_k$ are the constants to be determined. The choice of the independent variable $x$ stems from the experience gained in our foregoing work \citep{kaw1991}.  \par
After some experimentation, we found that $K_1=8$ yields a satisfactory fit to the reference values.
The polynomial approximation formula thus obtained is as follows.
 \begin{eqnarray}
  H_{\rm app}(1,\mu)&=&9.999982706853756\times 10^{-1}+ 3.465443224211651\times 10^{-4}x \nonumber \\
& & -1.411107006687451\times 10^{-2}x^2+ 3.269177042230116\times 10^{-1}x^3 \nonumber \\
& &  +4.133809356648527 x^4-7.188546622876579 x^5+7.772939980710241 x^6 \nonumber \\
& &-3.883055730606847 x^7 +7.595128286312914\times 10^{-1} x^8.  \label{eq-Happ}
 \end{eqnarray}
\indent A comparison between the values of $H_{\rm app}(1,\mu)$ generated by the present formula and those obtained with the integral representation $H_{\rm integ}(1,\mu)$ \citep[see also] [Table 1]{dav2008} indicates that the maximum relative error is $2\times 10^{-6}$. The approximate values are found to be correct to at least the fifth decimal place, and even the figure at the sixth decimal place differs 
from the correct one by no more than one unit.
 \subsubsection{Approximate Formula for Non-Conservative Scattering}
For a set of $L$ equally spaced values of $\varpi_0$, we
compute the values of $H_{\rm app}(1, \mu)/H_{\rm integ}(\varpi_0,\mu)$ for $M$ values of $\mu$, which are taken to coincide with the Chebyshev collocation points to apply the Chebyshev polynomial approximation method described in \citet[]{pre1992}. Then, they are approximated by a $K_2$-th degree polynomial of $x (\equiv \mu^{1/4})$, as before:
\begin{equation}
H_{\rm app}(1, \mu)/H_{\rm integ}(\varpi_0,\mu)=1+\sum_{k=0}^{K_2} C_k(\varpi_0)x^k.    \label{H1byHpi0}
\end{equation} 
\indent For a given value of $\varpi_0 $, the values of the coefficients $C_k(\varpi_0)\quad (k=0,\cdots, K_2)$ are determined by 
fitting the right-hand side of Eq.(\ref{H1byHpi0}) to the values of $H_{\rm app}(1,\mu)/H_{\rm integ}(\varpi_0,\mu)$ computed at $M$ discrete points of $\mu$, as indicated above.
As a result, we obtain $K_2$ sets of $L$ values of $C_k(\varpi_0)$.  \par
For actual computations, we adopted $M=3,500$ and $L=10,001$ so that the value of $\varpi_0$ ran from zero to unity with a step size of $10^{-4}$.  
For each $k$, the 10,001 values of $C_k(\varpi_0)$ obtained with the Chebyshev polynomial approximations were finally approximated by an $N_{\rm P}$-th degree polynomial of $\sqrt{1-\varpi_0}$:
\begin{equation}
C_k(\varpi_0)=\sum_{n=0}^{N_{\rm P}}B_{k,n}\eta^n,  \label{eq-CB}
\end{equation}
where $\eta\equiv\sqrt{1-\varpi_0}$ and the coefficients $B_{k,n}$ were determined by the standard least square method while varying the value of $N_{\rm P}$ from 4 through 9. The use of $\sqrt{1-\varpi_0}$ as the independent variable is based on a premise 
drawn from another past work on the $H$-function for isotropic scattering \citep[]{kaw1992}.\par
To assess the quality of the resulting formula, we computed 
\begin{equation}
\displaystyle{H_{\rm app}(\varpi_0,\mu)=H_{\rm app}(1,\mu)/\left(1+\sum_{k=0}^{K_2}C_k(\varpi_0) x^k\right)  }
\end{equation}
at the same $48\times 21$ grid points on the $(\varpi_0,\mu)$-plane as used in Section 2.2, which were 
then compared with those of $H_{\rm integ}(\varpi_0,\mu)$. The best formula was found with $K_2=8$ and $N_{\rm P}=8$, whose coefficients $C_k(\varpi_0)\quad (k=1,\cdots, K_2)$ are shown below.
\allowdisplaybreaks
\begin{mathletters}
\begin{eqnarray}
C_0&=&
      -1.368687418901498\times 10^{-6}+6.744526217097578\times 10^{-5}\eta  \nonumber \\
    & & -8.816747094601710\times 10^{-4}\eta^2+ 4.731152489223286\times 10^{-3}\eta^3\nonumber \\
    & &  -1.352739541743824\times 10^{-2}\eta^4+ 2.236433018731980\times 10^{-2}\eta^5\nonumber \\
     & & -2.147081702708310\times 10^{-2}\eta^6+1.112257595951489\times 10^{-2}\eta^7 \nonumber \\
     & & -2.406003988429531\times 10^{-3}\eta^8\\
C_1&=&
       \phantom{+}8.737822937355147\times 10^{-5}\phantom{+} -5.250514244222347\times 10^{-3}\eta  \nonumber \\
    & &  +7.644952859355422\times 10^{-2}\eta^2 -4.664908220536214\times 10^{-1}\eta^3 \nonumber \\
     & &  +1.482688198325839\eta^4 -2.663033364728811\eta^5 \nonumber \\
     & &  +2.727252555244034\eta^6 -1.485444888951274\eta^7 \nonumber \\
     & &  +3.340921510758153\times 10^{-1}\eta^8 \\    
C_2&=&
     -1.427222952750036\times 10^{-3}+9.300028322140796\times 10^{-2}\eta \nonumber \\
 & & -1.413069914567426\eta^2+ 8.880428860986575\eta^3  \nonumber \\
 & & -2.866825946137678\times 10^{+1}\eta^4+5.178036196746675\times 10^{+1}\eta^5 \nonumber \\
 & & -5.307180734532348\times 10^{+1}\eta^6+2.885782084328829\times 10^{+1}\eta^7 \nonumber \\
 & & -6.471219440031649\eta ^8\\
C_3&=&  \phantom{+}9.066801756884433\times 10^{-3} -6.354984995808299\times 10^{-1}\eta \nonumber \\
 & &+1.021262226727643\times 10^{+1}\eta^2 -6.444360574298017\times 10^{+1}\eta^3 \nonumber \\
& &+2.105330190640368\times 10^{+2}\eta^4 -3.824039368443171\times 10^{+2}\eta^5 \nonumber \\
& &+3.930240665640704\times 10^{+2}\eta^6 -2.139686267143788\times 10^{+2}\eta^7 \nonumber \\
& &+4.800025272319539\times 10^{+1}\eta^8 \\    
C_4&=& -2.855922558150419\times 10^{-2}+3.880224653851042\eta \nonumber \\
 & & -3.174231079700075\times 10^{+1}\eta^2+2.303877926374539\times 10^{+2}\eta^3 \nonumber \\
 & & -7.626655021168267\times 10^{+2}\eta^4+1.394034249890738\times 10^{+3}\eta^5 \nonumber \\
& & -1.438476211044276\times 10^{+3}\eta^6+7.852393856327993\times 10^{+2}\eta^7 \nonumber \\
 & & -1.764969590005163\times 10^{+2}\eta^8 \\
C_5&=&  \phantom{+}4.941209676842531\times 10^{-2}-3.976393849244121\eta \nonumber \\
   & &+6.000178277203062\times 10^{+1}\eta^2-4.542543148444882\times 10^{+2}\eta^3 \nonumber \\
   & &+1.512146625692455\times 10^{+3}\eta^4 -2.779737284749243\times 10^{+3}\eta^5 \nonumber \\
   & &+2.880598698878311\times 10^{+3}\eta^6 -1.577451021926768\times 10^{+3}\eta^7 \nonumber \\
    & &+3.554375808436865\times 10^{+2}\eta^8  \\
C_6&=& -4.798519468590785\times 10^{-2}+4.112841572654386\eta \nonumber \\
 & & -6.655808348671680\times 10^{+1}\eta^2+5.000349699512032\times 10^{+2}\eta^3 \nonumber \\
 & &-1.672172432180451\times 10^{+3}\eta^4+3.091851778649070\times 10^{+3}\eta^5 \nonumber \\
 & &-3.218110914157008\times 10^{+3}\eta^6+1.768094273655673\times 10^{+3}\eta^7 \nonumber \\
  & & -3.994358424590589\times 10^{+2}\eta^8  \\
 C_7&=& \phantom{+}2.461700902387896\times 10^{-2} -2.233648393380449\eta \nonumber \\
  & &+3.900465646584139\times 10^{+1}\eta^2 -2.880699974056035\times 10^{+2}\eta^3 \nonumber \\
  & &+9.688954523412610\times 10^{+2}\eta^4-1.802235503900686\times 10^{+3}\eta^5 \nonumber \\
 & &+1.883990440310628\times 10^{+3}\eta^6-1.038462482861755\times 10^{+3}\eta^7 \nonumber \\
 & &+2.352061082130820\times 10^{+2}\eta^8 \\
 C_8&=&-5.211353622987505\times 10^{-3}+4.967427514273564\times 10^{-1}\eta \nonumber \\
  & &-9.292147966163522\eta^2+6.773895398390997\times 10^{+1}\eta^3 \nonumber \\
  & &-2.294206635762768\times 10^{+2}\eta^4+4.292903843888321\times 10^{+2}\eta^5 \nonumber \\
  & &-4.506396634901928\times 10^{+2}\eta^6+2.491623632369491\times 10^{+2}\eta^7 \nonumber \\
  & &-5.657192709351447\times 10^{+1}\eta^8
 \end{eqnarray}   
\end{mathletters} \par
The maximum relative error of the approximate values $H_{\rm app}(\varpi_0,\mu)$ given by this
formula on the grid points of $(\varpi_0,\mu)$ is $2.1\times 10^{-6}$ and occurs at $\varpi_0=0.996$ and $\mu=0$, whereas the mean of the relative errors is $3\times 10^{-7}$. The differences between the formula values and those of $H_{\rm integ}(\varpi_0,\mu)$ 
are within $\pm 4$ at the sixth decimal place. 
If rounded at the fifth decimal place, they agree with each other to the last digits  
except for the three cases with $(\varpi_0,\mu)=(0.9, 0.3), (0.995, 0.95), (0.999,0.7)$, and $(0.9998, 0.4)$, where the figure in the fourth decimal place differs from the correct number by one unit. The maximum relative error of the computed values for the zeroth moment $\alpha_0$ is $1.64\times 10^{-7}$, arising 
at $\varpi_0=1$, where we have 2.0000003285 instead of 2 by means of the 350th-degree Gauss-Legendre quadrature.  
\section{Conclusion}
We have developed a numerical procedure to evaluate with 11-figure accuracy the value of Chandrasekhar's $H$-function for
isotropic scattering for arbitrary sets of the single scattering albedo, $\varpi_0$, and the cosine of the zenith angle of the emergent or incident direction of radiation, $\mu$, using the integral form expression. This should prove useful not only for checking the accuracy of numerical values of $H(\varpi_0,\mu)$ computed by other types of approximations but also as a practical computational tool in applications.    \par
The rational approximation formula constructed 
on the basis of the reference data computed with the integral-form representation for $H(\varpi_0,\mu)$ 
is significantly more accurate than any available in the literature, which may well make up for the fact that it is longer than others.   
\acknowledgments
\noindent{\it Acknowledgments}\ \  We are grateful to the anonymous referee for the very constructive 
comments and letting us be aware of the useful integral representation of the $H$-function which we inadvertently overlooked.  SSL acknowledges support from NASA Grant NNX09AE85G.
\appendix
\section{$I_1$ up to the $\varepsilon^5$ Term}
Expanding the left-hand side of Eq.(\ref{eq-expand}) in a series in $x$ and keeping the terms up to and including the order of $x^4$, and analytically carrying out the integration of Eq.(\ref{eq-I1-integ}),
 we get
\begin{eqnarray}
I_1&\simeq&(2\ln\,\varepsilon-2-\ln\,3)\varepsilon+{1\over 45} \left[30(1-\mu^2)\ln\,\varepsilon+5\mu^2(2+3\ln\,3)-3(3+5\ln\,3)\right]\varepsilon^3  \nonumber \\
& &-\left[{617\over 15750}+{2\ln\,3\over 15}-{1\over 75}\mu^2(9+25\ln\,3)+{1\over 25}\mu^4(2+5\ln\,3)\right.  \nonumber \\
& &\left.-{2\over 15}(2-5\mu^2+3\mu^4)\ln\,\varepsilon\right]\varepsilon^5+o[\varepsilon]^6
 \end{eqnarray}
\indent If $\varepsilon=10^{-3}$ as in our numerical computations, $I_1=-1.691412801800870\times 10^{-2}$ for $\mu=0$, $ -1.691412671960849\times 10^{-2}$ for $\mu=1/2$, and $-1.691412282441017\times 10^{-2}$ for $\mu=1$ in contrast to  
$-1.691412801800667\times 10^{-2}$,  $-1.691412671960754\times 10^{-2}$, and $-1.691412282441016\times 10^{-2}$
  respectively 
obtained from Eq.(\ref{eq-I1}).  This fact indicates that the latter expression for $I_1$  is of sufficient accuracy for our present purposes.
\section{Alternative Integral Representation for Conservative Scattering}
In the case of $\varpi_0=1$, we have an interesting alternative to Eq.(3) 
 \citep[][]{das2007}:
\begin{equation}
H(1,\mu)=\sqrt{3}(1+\mu)\exp\left [-\int_{0}^{1}\theta(x)\frac{{\rm d}x}{x+\mu}\right],    \label{B1}
\end{equation}
where 
\begin{equation}
\displaystyle{\theta (x)={1\over \pi}{\rm atan2}\left[{\pi \over 2}x,1-{1\over 2}x\,\ln{1+x\over 1-x}\right]}. 
\end{equation}
and
\begin{equation}
{\rm atan2}\,(y,x)=\begin{cases}
                          \   \arctan\,(y/x)  &  x>0 \\
                          \   \pi+\arctan\,(y/x) & y\ge 0, \ x<0 \\
                          \   -\pi+\arctan\,(y/x) &  y<0, \ x<0  \\
                          \   \pi/2   &  y>0, \ x=0  \\
                          \  -\pi/2  & y<0, \ x=0  \\
                          \  {\rm undefined}  & y=0, \ x=0
                             \end{cases}
\end{equation}\par 
Some remarks  on Eq.(B1) must therefore be in order  in comparison with Eq.(3):
first of all, the integrand $\theta(x)/(x+\mu)$ involved in Eq.(B1) exhibits a very steep increase as we come close to  $x=1$ and    converges  to  a  limiting 
 value of   $1/(1+\mu)$  at $x=1$.  Furthermore, in the vicinity of 
 $x=0$, it also rises almost vertically from 0 to nearly 0.5 over a short range  in $x$
 for small but non-zero values of $\mu$.  This fact makes the numerical integration of this integrand   extremely difficult particularly if  $\mu$ is small.\par
For numerical tests of the efficiency of Eq.(\ref{B1}), we have widely varied  the  degree $N$ of the Gauss-Legendre quadrature employed for carrying out the required integration. The  resulting values of  $H(\mu)$   are shown in Table 1 for $N=300$, 3000, and 30000   in comparison with those obtained from Eq.(3) with $N=300$ coupled with Eq.(4) .   Also shown in the column designated by ''Das-Bera'' are the computational values taken from Table-8 of \citet{das2007}, who used the Simpson's one third rule (the number of division points is unknown). \par
 It is now obvious  that we need to employ  the Gauss-Legendre quadrature with $N\ge 30000$ to get the values of $H(\mu)$  by means of Eq.(\ref{B1}) correct down to the 10th decimal place, and that Eq.(3) together with the procedure discussed  in the text of the present work, requiring only $N=300$ or less , is significantly faster  at least for the purpose of generating  reference numerical values.  
\begin{table}[h]
\caption{Accuracy comparison of the values of $H(\mu)$ for conservative scattering }
\begin{tabular}{c|c|c|ccc}
\hline\hline
$\mu$&   Eq.(3)  &  Das-Bera      &   $N=300$    &   $N=3000$     &    $N=30000$ \\
\hline
 0.00 & 1.0000000000 & 1.0000000000
& 1.0000000446 & 1.0000000002 & 1.0000000000 \\
 0.05 & 1.1365748468 &1.1365748417
& 1.1365748951 & 1.1365748471 & 1.1365748468 \\
 0.10 & 1.2473504425 & 1.2473504371
& 1.2473504930 & 1.2473504428 & 1.2473504425 \\
 0.20 & 1.4503514128 & 1.4503514071
& 1.4503514667 & 1.4503514131 & 1.4503514128 \\
 0.30 & 1.6425222645 & 1.6425222585
& 1.6425223208 & 1.6425222648 & 1.6425222645 \\
 0.40 & 1.8292756032 & 1.8292755970
& 1.8292756614 & 1.8292756035 & 1.8292756032 \\
 0.50 & 2.0127787700 & 2.0127787636
& 2.0127788298 & 2.0127787703 & 2.0127787700 \\
 0.60 & 2.1941330193 & 2.1941330128
& 2.1941330805 & 2.1941330197 & 2.1941330193 \\
 0.70 & 2.3739749125 & 2.3739749059
& 2.3739749748 & 2.3739749129 & 2.3739749125 \\
 0.80 & 2.5527043168 & 2.5527043101
& 2.5527043801 & 2.5527043172 & 2.5527043168 \\
 0.90 & 2.7305876649 & 2.7305876581
& 2.7305877289 & 2.7305876652 & 2.7305876649 \\
 1.00 & 2.9078105291 & 2.9078105222
& 2.9078105939 & 2.9078105294 & 2.9078105291 \\
\hline
\end{tabular}
\end{table}




\begin{thebibliography}{}
\bibitem[Bosma and de Rooij (1983)]{bos1983}Bosma, P.B., \& de Rooij, W.A. 1983,  \aap, 126, 283
\bibitem[Chandrasekhar (1960)]{chan1960}Chandrasekhar, S. 1960, " Radiative Transfer," Dover Publications, Inc., New York  
\bibitem[Das and Bera (2007)]{das2007}Das, R.N., \& Bera, R. 2007,  arXiv: 0711.3336 [astro-ph]
\bibitem[Davidovi\'{c} et al. (2008)]{dav2008}Davidovi\'{c}, D.M., Vukani\'{c}, J., \& Arsenovi\'{c}, D. 2008, \icarus, 194, 389 
\bibitem[Domke (1988)]{dom1988}Domke, H. 1988, \jqsrt, 39, 283 
\bibitem[Hapke (1993)]{hap1993}Hapke, B. 1993, \textit{Theory of Reflectance and Emittance Spectroscopy} (New York: \\
Cambridge University Press) 
\bibitem[Karanjai and Karanjai (1991)]{kar1991}Karanjai, S., \& Karanjai, M. 1991, \apss, 178, 331
\bibitem[Karanjai and Sen (1971)]{kar1971}Karanjai, S., \& Sen, M. 1971, \apss, 13, 267
\bibitem[Kawabata et al. (1991)]{kaw1991}Kawabata, K., Satoh, T., \& Ueno, S. 1991, \apss, 182, 249 
\bibitem[Kawabata et al. (1992)]{kaw1992}Kawabata, K. \& Satoh, T. 1992, \jqsrt, 47, 1 
\bibitem[Press et al. (1992)]{pre1992} Press, W.H., Teukolsky, S.A., Vettering, W.T., \& Flannery, B.P. 1992, \\ 
\textit{Numerical Recipes in FORTRAN}, 2nd edition (New York: \\
Cambridge University Press)
\bibitem[Rutily and Bergeat (1987)]{Rut1987}Rutily, B., \& Bergeat, J. 1987, \jqsrt, 38, 47  
\end{thebibliography}
\end{document}